# Superluminal Behaviors of Modified Bessel Waves*

WANG Zhi-Yong**, XIONG Cai-Dong

*School of Physical Electronics, University of Electronic Science and Technology of China, Chengdu 610054*

*Much experimental evidence of superluminal phenomena has been available by electromagnetic wave propagation experiments, with the results showing that the phase time do describe the barrier traversal time. Based on the extrapolated phase time approach and numerical methods, we show that, in contrary to the ordinary Bessel waves of real argument, the group velocities of modified Bessel waves are superluminal, and obtain the following results: 1) the group velocities increase with the increase of propagation distance, which is similar to the evanescent plane-wave cases; 2) for large wave numbers, the group velocities fall off as the wave numbers increase, which is similar to the evanescent plane-wave cases; 3) for small wave numbers, the group velocities increase with the increase of wave numbers, this is different from the evanescent plane-wave cases.*

The motion of a Bessel wave is of great interest in physics [1-4]. Whether an ordinary Bessel wave, NOT through a tunneling region, has also the genuine superluminal behaviors, this is a controversial issue: both affirmative and opposed viewpoints have been published in some literatures [5-9]. Nevertheless, contrary to the ordinary Bessel waves of real argument, we shall show that modified Bessel waves, described by the Bessel functions of imaginary argument, do have the superluminal behaviors. To avoid confusing an ordinary Bessel wave through a tunneling region with the modified Bessel wave, in especial, to avoid mixing our work with those literatures dealing with the passage of a Bessel wave through a tunneling region [10-11], let us point out that:

1) The ordinary Bessel waves via which the superluminal phenomena have been studied before [5-11], can be regarded as a cylinder which has no end plates and is open on both ends, have a plane wavefront which is perpendicular to the symmetry axis of the cylinder, and hence they propagate along the axial direction of the cylinder; 2) The Bessel waves studied in this paper correspond to the ones that propagate along the radial direction of a cylinder which has no lateral face and is open on circumference, with a cylindrical wavefront being perpendicular to the radial direction. For our purpose, we shall deal with the passage of such Bessel waves through a tunneling region, for the moment these Bessel waves appear as the so-called modified Bessel waves.

In Cartesian coordinate system $(x, y, z)$, assume that a hollow metal cylinder of radius $\rho \to +\infty$ has its axis coincident with the $z$-axis, two perfectly conducting plates of radius $\rho \to +\infty$ are $L$ apart, they form two parallel end plates of the cylinder and are respectively located at $z=\pm L/2$. The cylinder has no lateral face. A linear antenna with length $d \leq L$ is placed at the symmetry axis of the cylinder extending from $z=-d/2$ to $z=d/2$, and $d$ is small compared to the wavelength $\lambda$ of the fields produced by the antenna. In the present case, the boundary condition suggests the use of cylindrical coordinate system $(r, \theta, z)$ ($x=r\cos\theta$, $y=r\sin\theta$). Let $\omega$ denote the angular frequency, $k=\omega/c$ the wave number, $c$ the velocity of light in vacuum. As we know, the fields in the far zone $kr \gg 1$ (where $r$ is the distance from the source to the field point) represent the radiation fields of the antenna, for the objectives of the present paper, we limit our attention to just the far-zone fields. Let $\varphi(r, \theta, z, t)$ denote the radiation field intensities of the antenna, in cylindrical coordinates $(r, \theta, z)$, the wave equation for $\varphi$ takes the form ($kr \gg 1$)

$$\frac{1}{r}\frac{\partial}{\partial r}(r\frac{\partial \varphi}{\partial r}) + \frac{1}{r^2}\frac{\partial^2 \varphi}{\partial \theta^2} + \frac{\partial^2 \varphi}{\partial z^2} - \frac{1}{c^2}\frac{\partial^2 \varphi}{\partial t^2} = 0. \quad (1)$$

Assume a separable solution $\varphi(r, \theta, z, t) = R(r)\Theta(\theta)Z(z)\exp(i\omega t)$, where $R(r)$ describes the propagation characteristic of the waves and we shall only discuss it. In the present case, the boundary conditions are: $R(+\infty) < +\infty$, $\Theta(0)=\Theta(2\pi)$,

*Work supported by the Doctoral Program Foundation of Institution of Higher Education of China (Grant No. 20050614022).
**Corresponding author. E-mail: zywang@uestc.edu.cn

$\partial \Theta(0)/\partial \theta = \partial \Theta(2\pi)/\partial \theta$ and $Z(\pm L/2)=0$. Furthermore, for our purpose, we will limit our attention to just the far-zone fields of the antenna, i.e., will only study the radiation fields that are emitted by the antenna placed at the axis of the cylinder and propagate radially from $r=r_0 \gg 1/k$ to $r=\rho \to +\infty$. Substitute these into Eq. (1), one can obtain ($k=\omega/c$)

$$\begin{cases} \dfrac{d^2\Theta(\theta)}{d\theta^2} + \eta\Theta(\theta) = 0 \\ \Theta(0) = \Theta(2\pi),\ \Theta'(0) = \Theta'(2\pi) \end{cases}, \quad (2)$$

$$\begin{cases} \dfrac{d^2 Z}{dz^2} + \alpha Z = 0 \\ Z(\pm L/2) = 0 \end{cases}, \quad (3)$$

$$\begin{cases} \dfrac{1}{r}\dfrac{d}{dr}(r\dfrac{dR}{dr}) + (k^2 - \alpha - \dfrac{\eta}{r^2})R = 0 \\ R(+\infty) < +\infty,\ 1/k \ll r_0 \le r < +\infty \end{cases}. \quad (4)$$

From Eq. (2) one has $\eta=m^2$ ($m=0,1,2\ldots$), and Eq. (3) gives $\alpha = (2n\pi/L)^2$ ($n=1,2,3\ldots$), then Eq. (4) becomes

$$\begin{cases} \dfrac{1}{r}\dfrac{d}{dr}(r\dfrac{dR}{dr}) + [k^2 - (\dfrac{2n\pi}{L})^2 - \dfrac{m^2}{r^2}]R = 0 \\ R(+\infty) < +\infty,\ 1/k \ll r_0 \le r < +\infty \end{cases}. \quad (5)$$

Let $k_n=2n\pi/L$ and $\sigma_n = \sqrt{k^2 - k_n^2}$, the radial equation (5) can be put in a standard form by the change of variable $r \to s = \sigma_n r$, i.e., the Bessel equation. In terms of the order-$m$ Bessel functions of the first, $J_m(x)$, and the order-$m$ Neumann functions, $Y_m(x)$, one can write the solutions of Eq. (5) as, generally

$$R_{mn}(r) = C_m J_m(\sigma_n r) + D_m Y_m(\sigma_n r), \quad (6)$$

with $C_m$ and $D_m$ constants. The general solution of Eq. (1) can be written as

$$\varphi(r,\theta,z,t) = \sum_{m,n}[R_{mn}(r)\Theta_m(\theta)Z_n(z)]\exp(i\omega t). \quad (7)$$

Obviously, as $k^2 \ge k_n^2 = (2n\pi/L)^2$, the radial solution $R(r)$ of Eq. (5) is described by the Bessel functions of real argument (i.e., the ordinary Bessel functions), and the radiation fields $\varphi(r, \theta, z, t)$ described by the ordinary Bessel waves represent the traveling waves that propagate along the radial direction of the cylinder, with a cylindrical wavefront perpendicular to the radial direction. On the other hand, as $k^2 < k_n^2 = (2n\pi/L)^2$, the radial solution $R(r)$ is described by the Bessel functions of imaginary argument (or the modified Bessel functions). In terms of the modified Bessel functions of the first and second kind (denoted as $I_m(x)$ and $K_m(x)$, respectively), the radial solution of Eq. (5) can be written as

$$R_{mn}(r) = E_m I_m(g_n r) + F_m K_m(g_n r), \quad (8)$$

where $g_n = \sqrt{k_n^2 - k^2} > 0$, $E_m$ and $F_m$ are constants. Owing to the boundary condition $R(+\infty)<+\infty$, the coefficients $E_m \equiv 0$, then one has

$$R_{mn}(r) = F_m K_m(g_n r). \quad (9)$$

The asymptotic form $K_m(x) \sim \sqrt{\pi/2x}\exp(-x)$ ($x \to +\infty$) implies that, as $k^2 < k_n^2 = (2n\pi/L)^2$, the radiation fields described by *the modified Bessel waves* correspond to evanescent waves.

In order to study the superluminal behaviors of modified Bessel waves, let us have recourse to the analogy between the behaviors of modified Bessel waves and the quantum tunneling of particles. For this we assume that the separation $L$ between the two parallel end plates of the cylinder satisfies

$$L = \begin{cases} L_0 > 2\pi/k,\ r_0 \le r \le r_1 \\ L_1 < 2\pi/k,\ r_1 \le r \le r_2 \\ L_0 > 2\pi/k,\ r_2 \le r < +\infty \end{cases}, \quad (10)$$

where $r_1 > r_0 \gg 1/k$ corresponds to the radiation region of the antenna. Consider that the radiation fields $\varphi(r, \theta, z, t)$ with wave numbers distributed sharply around the given $k$ propagate along the radial direction, and because of the axial symmetry, we limit our attention to just the radial solution $R(r)$ and the time factor $\exp(i\omega t)$ is omitted. For simplicity, let $m=0$ and $n=1$ without loss of generality, then Eq. (5) becomes

$$\begin{cases} \dfrac{1}{r}\dfrac{d}{dr}(r\dfrac{dR}{dr}) + h^2 R = 0 \\ R(+\infty) < +\infty \end{cases}, \quad (r > r_0 \gg 1/k), \quad (11)$$

where

$$h = \begin{cases} h_0 = \sqrt{k^2 - (2\pi/L_0)^2},\ r_0 \le r \le r_1 \\ h_1 = i\sqrt{(2\pi/L_1)^2 - k^2} \equiv i\kappa,\ r_1 \le r \le r_2 \\ h_0 = \sqrt{k^2 - (2\pi/L_0)^2},\ r_2 \le r < +\infty \end{cases}. \quad (12)$$

Owing to Eq. (10), one has $h_0>0$ and $\kappa>0$. In terms of the Hankel functions of the first and second kind, $H_m^{(1)}(x) \equiv J_m(x) + iY_m(x)$, $H_m^{(2)}(x) \equiv J_m(x) - iY_m(x)$, as well as the modified Bessel functions of the first and second kind, the solutions of Eq. (11) can be written as ($m=0$ and $n=1$)

$$R(r) = \begin{cases} R_1 = a_1 H_0^{(1)}(h_0 r) + a_2 H_0^{(2)}(h_0 r), & r_0 \leq r < r_1 \\ R_2 = b_1 I_0(\kappa r) + b_2 K_0(\kappa r), & r_1 \leq r < r_2 \\ R_3 = \eta H_0^{(1)}(h_0 r), & r_2 \leq r < +\infty \end{cases}$$

(13)

The five unknown quantities $a_1$, $a_2$, $b_1$, $b_2$ and $\eta$ are restricted by the equations that impose continuity of $R(r)$ and $dR(r)/dr$ at $r=r_1$ and $r=r_2$. The quantity $H_0^{(1)}(h_0 r)$ represents the incident wave propagating from $r=r_0 \gg 1/k$ to $r=r_1$, while $H_0^{(2)}(h_0 r)$ represents the reflected wave that is reflected at $r=r_1$. Therefore the quantity $|a_2|^2 / |a_1|^2$ represents the reflection probability, and the transmission probability is $|\eta|^2 / |a_1|^2$. Likewise, the quantities $K_0(\kappa r)$ and $I_0(\kappa r)$ are the evanescent and anti-evanescent waves, respectively. Therefore, the annular region between $r=r_1$ and $r=r_2$ plays the role of photonic potential barrier. Applying the continuity of $R(r)$ and $dR(r)/dr$ at $r=r_1$ and $r=r_2$, one can shows that the transmission coefficient $T$ is

$$T \equiv \frac{\eta}{a_1} = \frac{pq}{a_{11}b_{11} + a_{12}b_{21}} \equiv |T|\exp[i\alpha(k)], \quad (14)$$

where $|T|$ stands for the amplitude or module of $T$, while the real function $\alpha(\kappa)$ represents the phase of $T$, and

$$\begin{cases} p = h_0[H_0^{(1)}(h_0 r_1) H_1^{(2)}(h_0 r_1) - H_0^{(2)}(h_0 r_1) H_1^{(1)}(h_0 r_1)] \\ q = \kappa[I_0(\kappa r_2) K_1(\kappa r_2) + I_1(\kappa r_2) K_0(\kappa r_2)] \\ a_{11} = h_0 H_1^{(2)}(h_0 r_1) I_0(\kappa r_1) + \kappa H_0^{(2)}(h_0 r_1) I_1(\kappa r_1) \\ a_{12} = h_0 H_1^{(2)}(h_0 r_1) K_0(\kappa r_1) - \kappa H_0^{(2)}(h_0 r_1) K_1(\kappa r_1) \\ b_{11} = \kappa K_1(\kappa r_2) H_0^{(1)}(h_0 r_2) - h_0 K_0(\kappa r_2) H_1^{(1)}(h_0 r_2) \\ b_{21} = \kappa I_1(\kappa r_2) H_0^{(1)}(h_0 r_2) + h_0 I_0(\kappa r_2) H_1^{(1)}(h_0 r_2) \end{cases}$$

(15)

Consider that analogous tunneling experiments with classical evanescent electromagnetic modes have shown the results are essentially in agreement with the quantum mechanical phase time approach [12] (note that the phase time has nothing to do with the phase velocity), we shall show the superluminal behaviors of modified Bessel waves via the phase time approach. To gain physics insight, we restrict ourselves to the extrapolated phase time approach [13-15] and, approximately, attribute to the extrapolated phase time the physics meaning of the transit time for the modified Bessel waves propagating through the region between $r=r_1$ and $r=r_2$. In our case, following the usual procedure, one can obtains the extrapolated phase time

$$\tau_T(r_1, r_2; k) = \left(\frac{d\omega}{dk}\right)^{-1}\left[r_2 - r_1 + \frac{d\alpha(k)}{dk}\right] \equiv \frac{r_2 - r_1}{v}, \quad (16)$$

where $d\omega/dk=c$ is the speed of light in vacuum, while $v$ represents the group velocity of the modified Bessel waves propagating through the region between $r=r_1$ and $r=r_2$ (i.e. the group velocity of the modified Bessel waves tunneling through photonic potential barrier). Using Eq. (14), one has

$$v = v(w,k) = \frac{cw}{w + \text{Im}[d(\ln T)/dk]}, \quad (17)$$

where Im($X$) represents the imaginary part of $X$ and $w=r_2-r_1$. Using Eqs.(16)-(17) we show the superluminal behavior of the modified Bessel waves via numerical methods (by *Mathematica* 5.0), and give the diagrammatic curves related to the ratio $v/c$ instead of the group velocity $v$ itself (set $c=1$). For simplicity let $h_0 \approx k$, $L_1=\pi/10$, $r_1=100$, then $r_2=100+w$ and $\kappa = \sqrt{400-k^2}$.

Following Figs.1-2 we summarize the superluminal behavior of the modified Bessel waves as follows:

(1) As shown in Fig.1, the group velocity $v$ of the modified Bessel waves is proportional to the width $w=r_2-r_1$, which accords with the so-called Hartman effect [16];

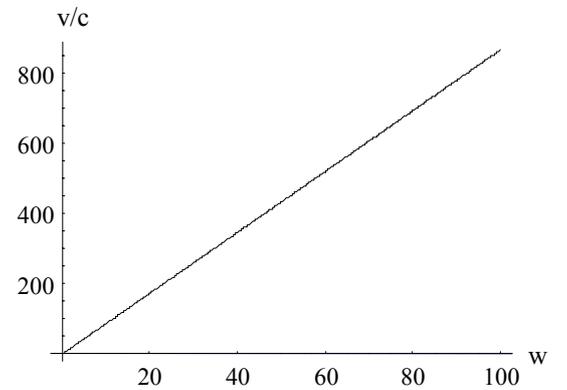

Fig.1 The group velocity $v$ as the function of the width $w$ ($k=10$ and $w=0\rightarrow100$). This graph is valid for both the modified Bessel waves and the evanescent plane-waves.

(2) As shown in Fig.2, for the wave number $k$ satisfying $k \geq k_0 \approx 0.6$, the group velocity $v$ of the modified Bessel waves decreases as the wave number increases, which implies that the group velocity $v$ increases as the height of photonic barrier increases;

(3) As shown in Fig.2, for $0.6 \approx k_0 \geq k > 0$, the group velocity $v$ rapidly increases as the wave number increases, which implies that the group velocity $v$ increases as the height of photonic barrier decreases.

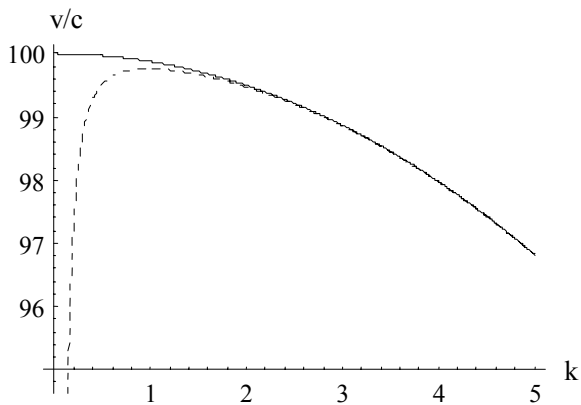

Fig.2 The group velocity v as the function of the wave number k (w=10 and k=0.01→5). The chain line stands for the modified Bessel waves, the solid line to the evanescent plane-waves.

Furthermore, to compare the superluminal behaviors of the modified Bessel waves with those of the evanescent plane-waves (the latter have been studied in photonic tunneling experiments), in Fig.1-2 we also give the diagrammatic curves for the evanescent plane-waves that tunneling through a photonic barrier (with the width $w=r_2-r_1$, and the photons within the photonic barrier satisfy $\kappa = \sqrt{400-k^2}$ ), from which one can show that they are similar to each other for the above statements (1) and (2), but different from each other for (3). That is, as $0.6 \approx k_0 \geq k > 0$, with the increase of wave number $k$, the group velocities of the modified Bessel waves rapidly increase while those of the evanescent plane-waves fall off. The behaviour of the group velocity for small wave numbers could be simply due to a violation of the underlying assumption that only the far field is regarded. For the given values of $L$ and $r$, $kr \gg 1$ is no longer true if $k$ is small, and near-field effects should be taken into account.

The authors would like to express his appreciation to Professor O. Keller for helpful correspondence and many interesting discussions on these matters, and would like to thank Professor G. Nimtz for his helpful comments.